# Demonstration of High-Gain Harmonic Lasing in a Terahertz Free-Electron Laser

## Author Information

Yin Kang[1], Cheng Yu[1], Yue Wang[2], Weiyi Yin[1], Zhangfeng Gao[1], Hanghua Xu[1], Hang Luo[1], Jian Chen[1], Taihe Lan[1], Xiaoqing Liu[1], Jinguo Wang[1], Huan Zhao[1], Fei Gao[1], Liping Sun[1], YanYan Zhu[1], Yongmei Wen[1], Chengcheng Xiao[1], Yongfang Liu[1], Yixuan Liu[1], Xingtao Wang[1], Jiaqiang Xu[1], Zheng Qi[1], Tao Liu[1], Bin Li[1, 3], Kaiqing Zhang[1, 3*], Zhen Wang[1*], Chao Feng[1, 3*], Bo Liu[1, 3] & Zhentang Zhao[1, 3]

## Affiliations

[1]Shanghai Advanced Research Institute, Chinese Academy of Sciences, Shanghai, China

[2]Zhangjiang Laboratory, Shanghai, China

[3]University of Chinese Academy of Sciences, Beijing, China

Correspondence to: Kaiqing Zhang, zhangkq@sari.ac.cn; Zhen Wang, wangz@sari.ac.cn; Chao Feng, fengc@sari.ac.cn

## Abstract

Compact Free-Electron Lasers (FELs) offering broad, continuous spectral tunability are traditionally constrained by fixed-parameter magnetic structures and the necessity for high-energy electron beams. High-gain Harmonic Lasing (HL) has long been proposed as a solution to overcome these limitations; however, a robust experimental verification of this principle has remained absent. Here, we report the first experimental demonstration of high-gain HL. By employing a frequency-tunable electron beam density modulation to dominate the fundamental instability, we achieved sustained FEL amplification at the 3$^{rd}$ and 5$^{th}$ harmonics of the wiggler. The HL mode generated output power comparable to conventional fundamental operation with enhanced stability and narrower spectral bandwidth. Notably, we demonstrate that HL extends the spectral coverage by a factor of two under fixed facility constraints, achieving pulse energies up to 540 µJ. These results establish high-gain HL as a versatile mechanism for advancing compact, wavelength-flexible FEL facilities.

# Introduction

Free-Electron Lasers (FELs) stand as one of the most significant advancements in photon science, capable of delivering coherent, high-intensity electromagnetic radiation across an unprecedented range, from the terahertz (THz) region to hard X-rays [1-4]. These revolutionary sources provide enabling tools for a broad spectrum of research fields, including ultrafast dynamics [5, 6], quantum material control [7, 8], and complex biological imaging [9, 10]. However, realizing the full potential of FEL technology, specifically, providing widely tunable and continuously available photon coverage while maintaining a manageable, compact facility footprint, is fundamentally restricted.

The key component enabling light emission in accelerator-based light sources is the magnetic insertion device (undulator or wiggler), which consists of a periodic array of magnets with alternating polarity. In an FEL employing a planar wiggler, the resonant wavelength $\lambda_n$ (and corresponding resonant frequency $f_n = \frac{c}{\lambda_n}$, where $c$ is the speed of light) is intrinsically governed by the relativistic energy of the electron beam $\gamma$, and the magnetic field strength parameter $K$, according to the resonance condition [11, 12]:

$$\lambda_n = \frac{\lambda_w}{2n\gamma^2}(1+K^2),$$

or

$$f_n = \frac{2cn\gamma^2}{\lambda_w(1+K^2)}, \ n = 1,3,5 \ldots \ldots, \ (1)$$

where $n$ is the harmonic number ($n = 1, 3, 5 \ldots \ldots$), $\lambda_w$ is the wiggler period, and $K = 0.934\lambda_w[cm]B[T]$, with $B$ representing the peak magnetic field strength. Conventionally, accessing shorter wavelengths or achieving broad tunability at the fundamental frequency ($n = 1$) demands either high electron beam energies (requiring large-scale accelerators), ultrashort wiggler periods, or operation in a low-$K$ regime, which often compromises FEL gain and stability. Consequently, achieving wide-range continuous tunability and facility compactness remains a critical challenge in the advancement of FEL technology.

To overcome these fundamental limitations, research efforts have long explored advanced FEL concepts, notably Harmonic Lasing (HL), where the FEL lasing at high harmonics of the wiggler [13-21]. The HL concept offers a pathway to extend coverage to shorter wavelengths without resorting to high beam energies, thereby enabling the development of more compact facilities with broader spectral tunability. In general nonlinear HL within a self-amplified spontaneous emission (SASE) FEL [22, 23], radiation at the fundamental wavelength dominates the interaction. Consequently, the harmonic radiation remains inherently orders of magnitude weaker (typically ~1% power for the $3^{rd}$ harmonic and ~0.1% for the $5^{th}$) [24-29].

The fundamental physics of high-gain HL lies in the suppression of the exponential gain of the dominant fundamental (n=1) instability to enable the sustained, high-efficiency exponential amplification of a desired higher odd harmonic. Several methods have been proposed to achieve this critical suppression. One strategy involves installing periodic phase shifters between wiggler segments, engineered to disrupt the fundamental ponderomotive interaction while remaining transparent to harmonic growth [13, 14]. Another approach relies on precisely tailoring the initial electron beam distribution, such that it carries no bunching at the fundamental frequency, but is modulated specifically at the odd harmonics of the wiggler [15]. Both techniques aim to extinguish the fundamental frequency oscillation, allowing the target harmonic to quickly enter its exponential gain stage. Despite the compelling nature of these theoretical mechanisms, which promise intrinsically superior performances, a robust experimental verification of this principle in a high-gain FEL amplifier, necessary to fully validate the underlying physics and unlocking the full utility of HL, has remained absent.

In this work, we present the first experimental verification of the high-gain HL principle. Rather than relying on mechanical phase shifters, our approach utilizes a high-quality, frequency-tunable electron beam density modulation (pre-bunching) initialized precisely at the target harmonic frequency. This robust initial harmonic seed effectively bypasses the slower-growing fundamental instability, satisfying the critical theoretical requirement of suppressing the fundamental mode via gain domination. Validated in the terahertz (THz) regime, we confirm that this HL operation achieves efficiencies comparable to fundamental lasing (FL), but with enhanced stability and narrower bandwidth consistent with a pure harmonic interaction. These results establish HL as a critical mechanism for advancing compact, wavelength-tunable FEL technology.

**Results**

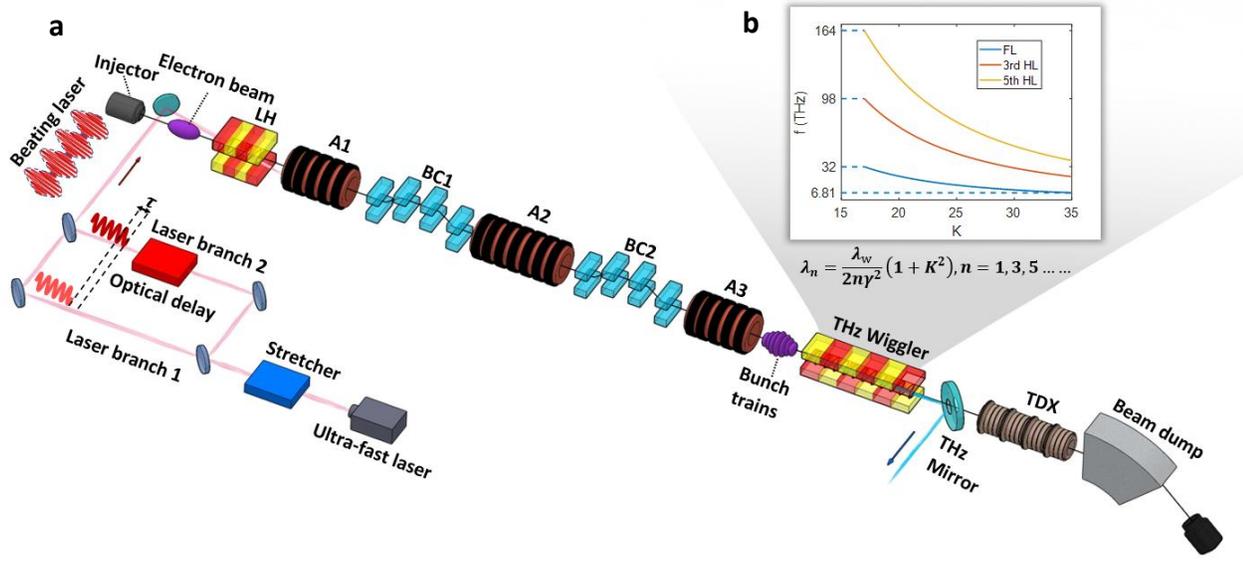

**Fig. 1: Experimental Setup and Harmonic Lasing (HL) Schematic. (a)** An energy-chirped electron beam produced from an injector, interacted with a frequency beating laser in the laser heater (LH), to acquire controlled energy modulation onto the beam. The energy modulated beam is converted into density modulation, and the modulated frequency is suppressed by two bunch compressors (BCs). The beam is sequentially accelerated to about 1GeV by several accelerator sections, bunch trains with tunable intervals are obtained by the end of the linac. And then the bunch trains pass through a wiggler to verify the high-gain HL principle. **(b)** The theoretical frequency coverages of FL, 3$^{rd}$ and 5$^{th}$ HL with a fixed *K* range (17-35) of the wiggler.

The experiment was conducted at the Shanghai soft-ray free-electron laser facility (SXFEL) [30, 31]. Fig. 1a illustrates the experimental setup and the schematic for HL (see methods for more details). The implementation of HL relies on controlled density modulation (pre-bunching) of the electron beam. The process begins with the energy modulation, where a 500-pC electron beam, produced by an injector with an energy of 115 MeV and a full bunch length of 14.3 ps, interacts with a laser beat-wave in the modulator. This interaction imparts a controlled energy modulation with a frequency $f_0$ tunable from 0.1 to 6 THz [31]. The beat-wave is generated by an 800 nm Ti: Sapphire laser source via optical heterodyning.

Subsequently, an energy-chirped is introduced and linearized by the accelerating section A1. The beam then passes through magnetic bunch compressors (BCs) to convert this energy modulation into precise, high-frequency density modulation. The modulation frequency is upshifted to $f_b = Cf_0$, where $C$ represents the total compression factor of the two BCs ($C \approx 10$ in this experiment). During this process, the modulation is enhanced by the collective effects within the accelerator [32, 33]. Simultaneously, the beam is accelerated to 1.08 GeV and the energy chirp is compensated by section A3. Finally, a beam with a full length of about 1 ps is obtained at the linac exit, where the phase space is characterized using an X-band transverse deflecting cavity (TDX).

Fig. 2 presents the characterization of these microbunching structures. The measured longitudinal phase space distributions and projected density profiles confirm the generation of electron bunch trains with tunable bunching frequencies, $f_b$, ranging from approximately 10 THz to 53 THz. This frequency control was achieved by adjusting the time delay $\tau$ between the beating laser pulses, corresponding to a path length difference of 0.7 mm to 8.4 mm.

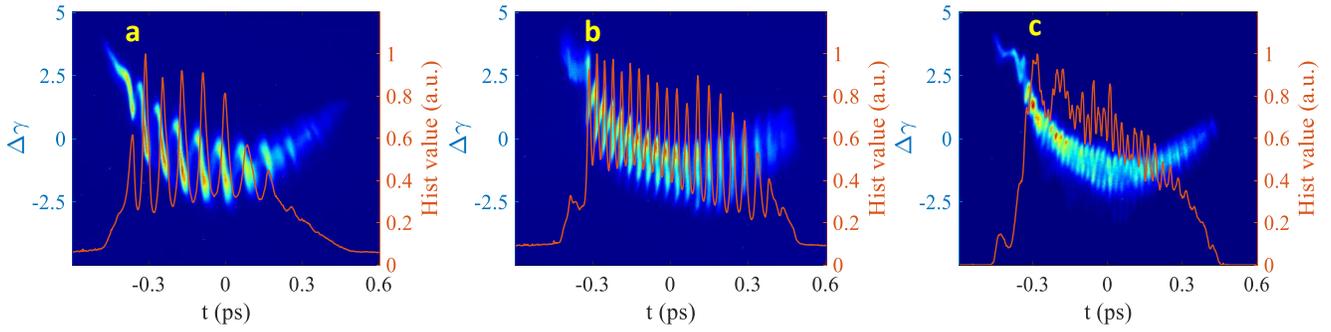

**Fig. 2: Measured Longitudinal Phase Space Distributions of the Electron Beam.** Panels display pre-bunched electron beams with bunching frequencies of 10 THz **(a)**, 25 THz **(b)**, and 53 THz **(c)**, by simply tuning the time delay of the beating laser $\tau$ from 0.7 mm to 8.4 mm. The orange lines give projected density profiles.

Following acceleration, the electron bunch trains are injected into a 5 m-long electromagnetic wiggler with a period of $\lambda_w = 0.28$ m ($N_w$=18) to generate FEL radiation at frequency $f_r$. As demonstrated in Ref [31], the FEL operates in the high-gain regime, delivering high peak intensity. The wiggler functions within an optimized $K$ parameter range of 17–35, a regime selected to ensure power supply stability and effective magnetic field correction. As illustrated in Fig. 1b, HL substantially extends the spectral coverage from the fundamental limit of 6.32–32 THz (FL) to 6.32–164 THz (using the 3$^{rd}$ and 5$^{th}$ harmonics). For HL operation, the fundamental resonance frequency of the wiggler $f_{wf}$ is detuned to exactly $1/n$ of the bunching frequency $f_b$ (i.e., $f_{wf} = f_b/n$). This ensures the HL frequency matches the pre-bunching frequency, effectively seeding the $n$-th harmonic of the wiggler prior to the onset of fundamental exponential growth. This dominant harmonic seed allows the harmonic interaction to reach the exponential gain regime much faster than the FL.

To demonstrate the performance of the HL mechanism, comparative experiments were conducted between HL and standard FL targeting an identical radiation frequency ($f_r = 32\ THz$). In the HL configuration, the wiggler was tuned to resonate at a subharmonic of the target frequency ($n = 3$, $f_{wf} = \frac{f_r}{3} = 10.6\ THz, K = 29.5$), characterized by a gain length of $L_{nh}$ and output pulse energy $E_{nh}$. In contrast, for the FL reference case, the wiggler was resonant directly at the target frequency ($n = 1$, $f_{wf} = f_r = 32\ THz$, $K = 17.0$), yielding a gain length $L_f$ and pulse energy $E_f$. Theoretical calculations using the experimental parameters predict near-unity ratios for gain length and pulse energy (see Methods), suggesting that both cases perform comparably at 32 THz. This similarity arises because, in the HL regime, the larger wiggler parameter $K$ and the increased radiation slippage length compensate for the inherently weaker higher-order interaction, thereby sustaining both the gain and the output pulse energy.

The comparison of the FEL performance is summarized in Fig. 3. Radiation properties, including pulse energy, spectrum, and transverse profile, were characterized by a dedicated THz diagnostic platform (see Methods). The autocorrelation functions of the electric field and corresponding spectra, measured by a Michelson interferometer, are shown in Fig. 3a and b. Compared to FL, the HL mode exhibits a longer pulse duration (Fig. 3a) and a slightly narrower bandwidth (4.35% vs. 5.76%, FWHM, Fig. 3b). This spectral narrowing and pulse lengthening validate the predicted dynamics of a pure harmonic interaction, governed by the $n$-times longer slippage length in the $n$-th harmonic HL mode. Time-dependent simulations (Fig. 3b, see Methods) show excellent agreement with these experimental results. Furthermore, the measured 3[rd] harmonic HL spectrum (Supplementary Fig. 1) reveals no Fourier component near the fundamental frequency, confirming that the fundamental mode has negligible effect on the HL process.

Transverse intensity profiles were measured using a THz camera after focusing with gold-coated off-axis parabolic mirrors and THz focusing lens. Fig. 3c and d display the results for FL and HL, respectively. The measured

FWHM dimensions of 53 *μm* (vertical) × 67 *μm* (horizontal) for FL and 55 *μm* (vertical) × 60 *μm* (horizontal) for HL indicate that the transverse mode quality is well preserved in the HL regime.

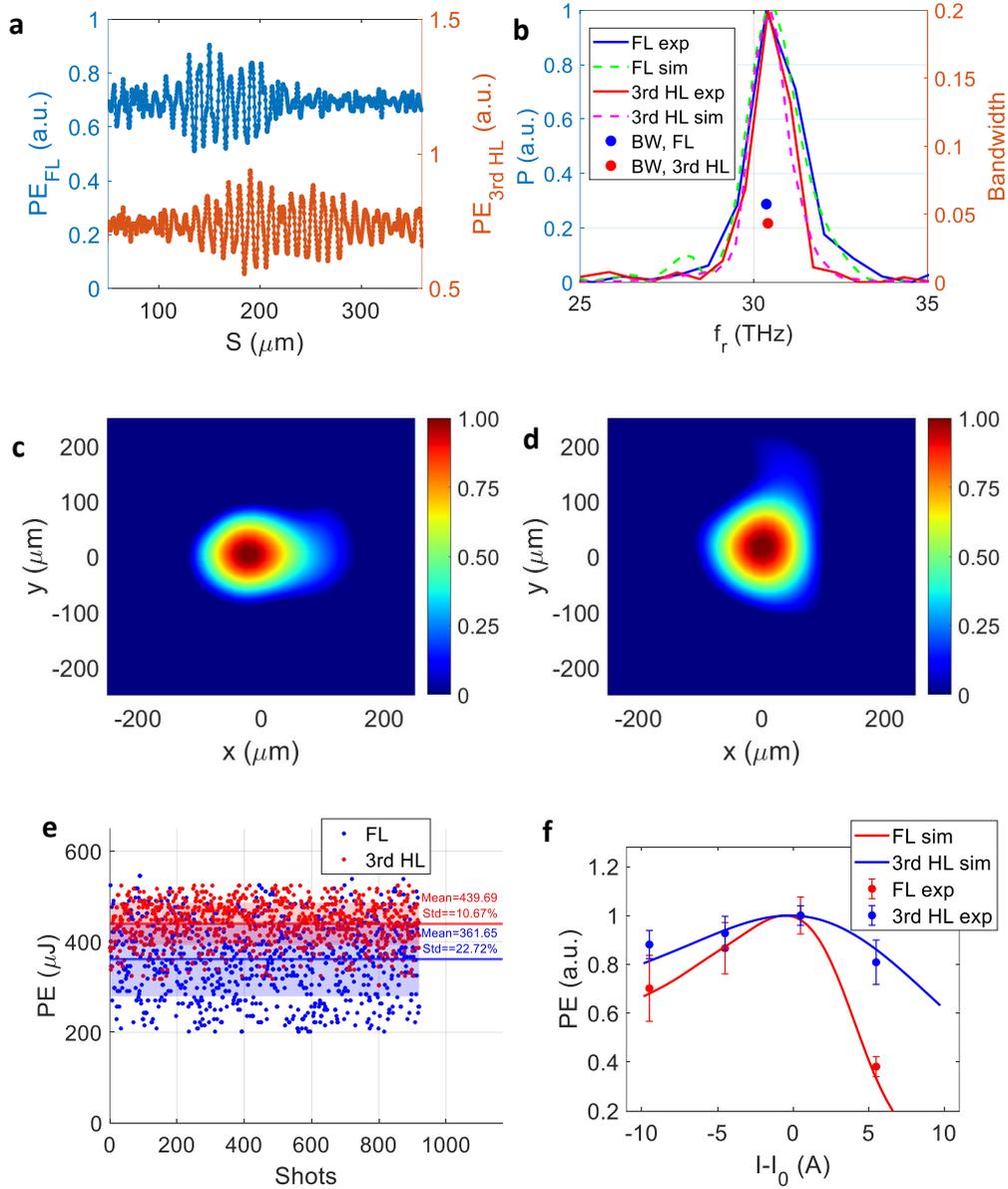

**Fig. 3: Comparation of FEL performances between FL and HL at the same frequency (32 THz).** Measured autocorrelation functions of the electric field **(a)** and corresponding spectra **(b)**. Transverse intensity profiles of the FL **(c)** and HL **(d)** modes. **(e)** Measured pulse energies for the fundamental and harmonic cases. **(f)** Measured and simulated pulse energies as a function of the wiggler power supply current detuning.

The THz pulse energies were measured using a calibrated Golay cell detector (background subtracted, see Methods) and the results are presented in Fig. 3e. Notably, the 3rd harmonic HL exhibits lower intensity fluctuation compared to FL (10.67% for HL vs. 22.7% for FL). This enhanced stability is attributed to the reduced sensitivity of the HL mode to power supply instabilities of the wiggler, a conclusion supported by the measurements in Fig. 3f and simulations in Supplementary Fig. 2. This robustness arises because the wiggler parameter $K$ becomes less sensitive to supply current variations in the high-current regime required for HL (see Supplementary Fig. 2 and 3). Consequently, the wiggler operating range is optimized in this large-$K$ region (17–35, corresponding to 110–300 A). In contrast, for FL, the same level of current jitter introduces an additional radiation power jitter of approximately 10% (Fig. 3f). For higher-frequency FL, this current sensitivity is expected to further degrade radiation stability.

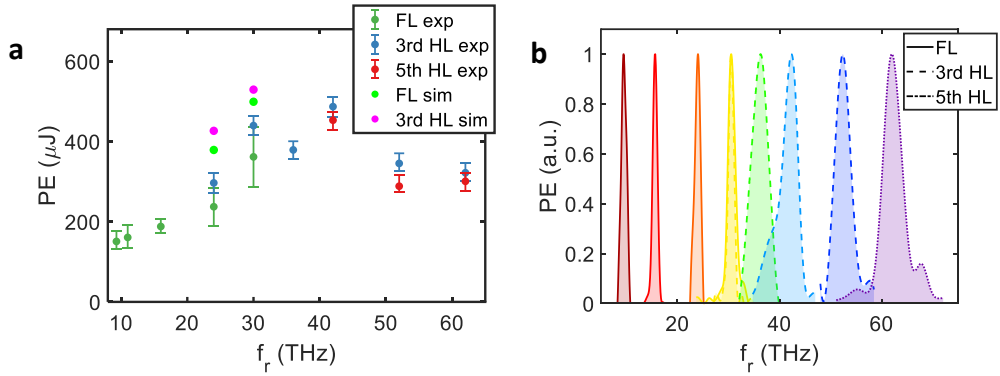

**Fig. 4: Spectral coverage extension using a limited Wiggler $K$ range.** Measured pulse energies **(a)** and spectra **(b)** for FL and 3rd, 5th HL.

The primary utility of this verified HL mechanism lies in its ability to extend continuous spectral coverage under fixed wiggler constraints. In our experimental setup, utilizing a fixed electron beam energy of 1.08 GeV and a $K$ parameter range of 17–35, FL was limited to a spectral coverage of 9–32 THz. However, by employing the HL mechanism (specifically the 3rd and 5th harmonics) within this identical $K$ range, the coverage was successfully extended to 9–62 THz. Fig. 4 summarizes the measured pulse energies and spectra. Notably, a maximum pulse energy of 540 μJ was achieved at 42 THz using the 3rd harmonic. This practical demonstration confirms that HL effectively overcomes the limitations imposed by a narrow $K$ tuning range, enabling substantial spectral broadening.

Furthermore, achieving pulse energies equivalent to those of the conventionally tuned fundamental mode provides direct experimental verification that the initial fundamental instability was successfully suppressed.

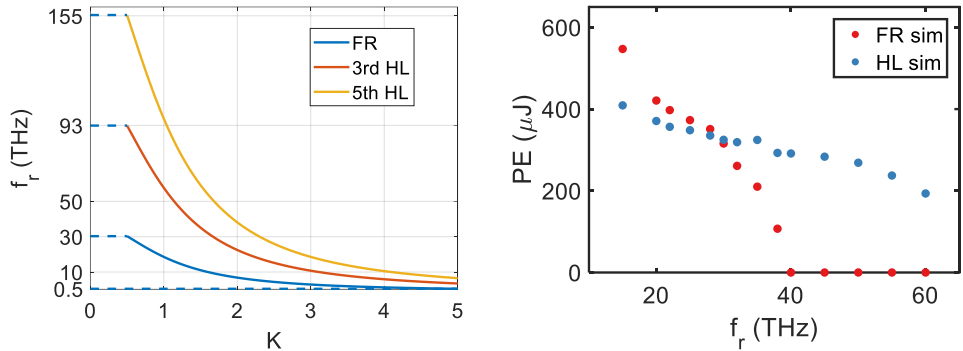

**Fig. 5: Calculated and simulated performance for a compact THz FEL. (a)** Relationship between radiation frequency $f_r$ and the resonant wiggler parameter $K$ for FL and HL (3rd and 5th harmonics). **(b)** Simulated saturated pulse energies across the spectral range of 5-60 THz.

Finally, we consider the application of this technique to a compact THz FEL. Simulations were performed using an electron beam with an energy of 20 MeV, a bunch charge of 2 nC, a normalized emittance of 1.5 μmrad and a bunch length of 2 ps, utilizing an undulator section with a period length of 2.35 cm and a total length of 10 m. Fig. 5a displays the output frequency coverage of this facility within a practical $K$ range. The results highlight that HL extends the spectral coverage from 0.5–30 THz (FL) to 0.5–155 THz. As shown in Fig. 5b, while FL is cut off above 40 THz as the required $K$ approaches zero, the 3rd harmonic HL maintains pulse energies exceeding 200 μJ up to 60 THz. These findings validate HL as an enabling technology for compact light sources, allowing access to shorter wavelengths without the need for high beam energies.

**Discussion**

In this work, we reported the first experimental demonstration of the high-gain HL principle in an FEL amplifier. By employing a pre-bunched electron beam, we achieved robust lasing at the 3rd and 5th harmonics. These results confirm that HL offers distinct advantages over fundamental operation. Primarily, it significantly broadens the accessible spectral range, reaching up to 62 THz in this work, under fixed facility parameters. Simultaneously, it enhances operational stability by allowing the wiggler to function in magnetically stable, high-*K* regimes, and offers the potential for improved spectral quality through narrower bandwidths due to the increased slippage length.

Although this demonstration utilized an electromagnetic wiggler in the THz regime, the implications extend to a wide range of FEL configurations. The observed stability is also beneficial for permanent magnet wigglers, where field quality typically degrades at the large gaps required for short-wavelength fundamental lasing. Moreover, HL emerges as an ideal strategy for compact, low-energy accelerators, such as superconducting industrial or university-scale sources [34]. By effectively decoupling the target wavelength from the strict electron beam energy constraints, HL enables these smaller facilities to access spectral domains previously restricted to large-scale facilities. Consequently, this mechanism, which is adaptable to various seeding schemes [3, 4, 35-39], promises to significantly enhance the versatility and reach of fully coherent photon sources.

## Acknowledgements

The authors thank Yu Zhang for helpful discussions. This work was supported by the National Natural Science Foundation of China under grant no. 12275340 (K. Z), 12105347 (K. Z) and 12435011 (C. F), CAS Project for Young Scientists in Basic Research under grant no. YSBR-115 (C. F), Shanghai Municipal science and Technology Major Project (C. F) and Innovation Program of Shanghai Advanced Research Institute, CAS under grant no. 2024CP001 (K. Z).

## Contributions

C. F, K. Z and Y. K conceived and designed the experiments. K. Z and Y. K conducted the experiments with the help from Z. W, Y. W (Yue Wang), Z. G, H. L , T. L (Tao Liu), Z. Q and with the software and hardware supports by C. Y, W. Y, H. X, J. C, X. L, T. L, J. W, H. Z, F. G, L. S, Y. Z, Y. W (Yongmei Wen), C. X., Y. L (Yongfang Liu), Y. L (Yixuan Liu), X. W, J. X, B. L (Bin Li). The simulations on beam dynamics were performed by Y. K. and Z. W.  The manuscript was written by C. F, K. Z and Y. K with contributions from Z. W. Management and oversight of the project was provided by C. F, B. L (Bo Liu) and Z. Z.

## Ethics declarations

Competing interests

The authors declare no competing interests.

## Methods

**Frequency-beating Laser System**

As shown in Fig. 1a, the frequency-beating laser system follows the layout in Ref [31]. To ensure effective modulation for final frequencies exceeding 30 THz, the system utilizes an initial laser pulse with an energy of up to 1 mJ. An 800 nm Ti:Sapphire laser source with a pulse length of 15 fs (FWHM) is first stretched to 20 ps by a laser stretcher. Then the pulse is split into two branches, one of which passes through an optical delay, before being recombined to form a tunable temporal intensity profile. This system generates a periodic field profile at the beating frequency $f_0$ ($f_0 = \frac{\mu\tau}{2\pi}$, where $\mu$ is the chirp rate of the laser pulse and $\tau$ is the time delay between two the branches). The beating frequency can be continuously adjusted from 0.1 to 6 THz by simply tuning the optical delay from 0.22 to 9 mm. Consequently, density modulation with frequencies ranging from 1-62 THz can be obtained after compression in the two magnetic bunch compressors (BC1 and BC2) within the linac.

**Accelerator Setup**

To achieve higher radiation pulse energy, a beam charge of 500 pC was used. The beam phase space was characterized by an X-band deflecting cavity, whose temporal resolution (~5 fs) was optimized to resolve the high-frequency bunching structures. Radiation was generated in a 5-meter-long electromagnetic wiggler (period 0.28 m, $N_u$=18), which can be tuned to the fundamental or a harmonic wavelength. For HL, the wiggler current was set to resonate at a subharmonic of the target frequency (Eq. 1, $n$ = 3, 5, 7…), requiring no other modifications to the setup. The current stabilities (110–270A) of wiggler are measured and shown in supplementary Fig. 3.

**THz diagnostic system**

The THz diagnostic platform, depicted in Fig. 1a and Supplementary Fig. 4, separates the radiation from the electron beam via a gold-coated mirror with a 2-mm aperture. The beam passes through the aperture, while the THz light is reflected through a 22-mm diamond window into the diagnostic line. Within this line, the light is directed and focused by a series of optics for specific measurements. Pulse energy is measured with a Golay cell (Golay cell-1, TYDEX, GC-1D), preceded by interchangeable band-pass or low-pass filters and switchable attenuators (1%, 10%, 30% transmission). Spatial profile measurements employ a movable mirror and THz focusing lens (Swiss THz optics F0.7) to focus the light onto a THz camera (Dolphin Optics, Beam 2000). For spectral analysis, a flip mirror redirects the light into a Michelson interferometer; the collimated beam is detected by a second Golay cell (Golay cell-2) to record the interferogram. Both Golay cells were calibrated using a continuous-wave $CO_2$ laser (10.6 μm wavelength), yielding responsivities of 1.83 μJ/V for Golay cell-1 and 3.29 μJ/m for Golay cell-2.

**Theoretical calculations and numerical simulations**

The gain length ratio and pulse energy ratio between HL and FL are given by [13]:

$$\frac{L_{nh}}{L_f} = \left(\frac{a_1 JJ_1(\xi_n)}{n^{\frac{1}{2}} a_n JJ_n(\xi_1)}\right)^{2/3}, \quad (2)$$

$$\frac{E_{nh}}{E_f} = \left(\frac{a_n JJ_1(\xi_n)}{a_1 JJ_n(\xi_1)}\right)^2 \frac{l_{nh}}{l_f}. \quad (3)$$

where $a_n$ is the wiggler parameter at $n$th harmonic, $JJ_n$ is the Bessel function factors associated with the wiggler, $\xi_n = a_n^2/2(1+a_n^2)$, $l_{nh}$ and $l_f$ is the pulse length of the HL and FL output, respectively. With the experimental parameters, $\frac{L_{nh}}{L_f}$ and $\frac{E_{nh}}{E_f}$ are calculated to be 1.15 and 1.03.

Start-to-end simulations were performed based on the measured electron beam parameters and the experimental setup. Beam dynamics from the photocathode to the LH were simulated with the particle-tracking code ASTRA [40], considering the LSC effects. The particle-tracking code ELEGANT [41], accounting for LSC and CSR effects, was used to simulate the laser modulation and beam dynamics from the LH to the entrance of THz wiggler. The FEL within the wiggler was simulated with the time-dependent code GENESIS [42].

## Data Availability

The data that support the findings of this study are presented in the paper and the Supplementary Information file. Other relevant data and findings of this study are available from the corresponding author upon reasonable request.